\documentclass[12pt,aps,nofootinbib]{revtex4-1}

\usepackage{amsmath,amssymb,bm}
\usepackage{graphicx}
\usepackage[breaklinks=true]{hyperref}
\usepackage{ascmac}
\usepackage{yfonts}
\usepackage{xcolor}

\newcommand{\re}{\mathrm{e}}
\newcommand{\rd}{\mathrm{d}}
\renewcommand{\d}{\partial}
\newcommand{\tr}{\mathrm{Tr}}

\newcommand{\vev}[1]{\left\langle #1 \right\rangle}

\newcommand{\er}[1]{Eq.~\eqref{#1}}

\newcommand{\sr}{\sqrt}
\newcommand{\fr}{\frac}

\newcommand{\lb}{\left}
\newcommand{\rb}{\right}

\newcommand{\U}{\mathrm{U}}
\newcommand{\SU}{\mathrm{SU}}

\DeclareMathOperator{\diag}{diag}

\begin{document}

\title{Quantum Hall liquids in high-density QCD}

\author{Kentaro Nishimura}
\affiliation{International Institute for Sustainability with Knotted Chiral Meta Matter (SKCM2),
Hiroshima University, Hiroshima 739-8511, Japan}
\author{Naoki Yamamoto}
\affiliation{Department of Physics, Keio University, Yokohama 223-8522, Japan}
\author{Ryo Yokokura}
\affiliation{Research and Education Center for Natural Sciences, Keio University, Yokohama 223-8521, Japan}

\begin{abstract}
There exist metastable domain walls of the flavor-singlet meson $\eta$ for the $\U(1)$ axial symmetry in two-flavor color superconductivity (2SC) in QCD at large baryon density. We show that, due to the coupling of $\eta$ to confined $\SU(2)$ gluons in the 2SC phase, the effective theory on the domain wall is described by the $\SU(2)_{-1}$ Chern-Simons theory, which is dual to the $\U(1)_{2}$ Chern-Simons theory. This theory has a spin-1 droplet excitation that does not carry a baryon number, which we identify as a vector meson.
We also discuss the effective theories and baryonic droplet excitations on the domain walls of the flavor-singlet mesons in the superfluid phases of QCD at large isospin density and two-color QCD at large baryon density.

\end{abstract}
\maketitle

\section{Introduction}
Quantum Hall liquid is a novel state of matter, described by the topological field theory, called Chern-Simons theory~\cite{Tong:2016kpv}. Its physical realization was originally found in condensed matter systems---two-dimensional electrons in a magnetic field. More recently, the relevance of the quantum Hall states in quantum chromodynamics (QCD) was pointed out in a seminal paper~\cite{Komargodski:2018odf} (see also, e.g., Refs.~\cite{Ma:2019xtx,Karasik:2020pwu,Kitano:2020evx,Bigazzi:2022luo} for related works): spin-$N_{\rm c}$/2 baryons in the QCD vacuum are interpreted as quantum Hall droplets in the large-$N_{\rm c}$ limit \cite{tHooft:1973alw,Witten:1979kh}, and such topological objects are described by the $\SU(N_{\rm c})_{-N_{\rm f}}$ Chern-Simons theory, which is dual to the $\U(N_{\rm f})_{N_{\rm c}}$ Chern-Simons theory. Here, $N_{\rm c}$ and  $N_{\rm f}$ are the numbers of colors and flavors, respectively. This should be contrasted with spin-$1/2$ baryons that are described as topological solitons known as skyrmions \cite{Skyrme:1961vq,Skyrme:1962vh}.

In this paper, we show that, in the two-flavor color superconductivity (2SC)~\cite{Bailin:1983bm,Alford:1997zt,Rapp:1997zu} in QCD at high density, there exist quantum Hall states described by the $\SU(2)_{-1}$ Chern-Simons theory, which is dual to the $\U(1)_{2}$ Chern-Simons theory.
% via the level-rank duality \cite{Hsin:2016blu}. 
Our argument is based on the effective theory on metastable domain walls of the flavor-singlet meson $\eta$ for the $\U(1)$ axial symmetry in the 2SC phase~\cite{Son:2000fh}.
The new essential ingredient is the topological coupling of $\eta$ to confined $\SU(2)$ gluons due to the $\SU(2)$ instanton effect in the effective theory on the $\eta$ domain wall. Here, $\eta$ can be treated as a light pseudo-Nambu-Goldstone mode within the framework of the low-energy effective theory, since the $\SU(3)$ instanton effect becomes weaker at higher density. 
Unlike the situation of the QCD vacuum~\cite{Komargodski:2018odf}, we need not take the large-$N_{\rm c}$ limit, and our description is under theoretical control for real $N_{\rm c}=3$ QCD {\it per se}.%
\footnote{Note that color superconductivity is suppressed in the large-$N_{\rm c}$ limit as the diquark pairing is not a color singlet~\cite{Deryagin:1992rw,Shuster:1999tn}. Therefore, the large-$N_{\rm c}$ scaling considered in Ref.~\cite{Komargodski:2018odf} does not apply to our setup.}
Moreover, while the quantum Hall droplet is realized as a spin-$N_{\rm c}/2$ baryon in the QCD vacuum~\cite{Komargodski:2018odf}, the droplet in our case is a spin-$1$ excitation that does not have a baryon number, which we identify as a vector meson.

We also study the effective theories on the domain walls of the flavor-singlet mesons in the superfluid phases of other dense QCD and QCD-like theories. One is QCD at large isospin density~\cite{Son:2000xc,Son:2000by}, and the other is two-color QCD at large baryon density~\cite{Kanazawa:2009ks}. We show that the effective theory on the domain wall in the former case is the $\SU(3)_{-2}$ Chern-Simons theory, which is dual to the $\U(2)_{3}$ Chern-Simons theory, while the latter is the $\SU(2)_{-2}$ Chern-Simons theory, which is dual to the $\U(2)_{2}$ Chern-Simons theory. We find that both theories have droplet excitations with the $\U(1)$ baryon number, which are identified as a spin-$3/2$ baryon and spin-$1$ baryon, respectively.

This paper is organized as follows. In Sec.~\ref{sec:2SC}, we review the known basic properties of the 2SC phase. In Sec.~\ref{sec:QHE}, we show the emergence of the quantum Hall liquids described by the Chern-Simons theory in the 2SC phase. We also discuss the spin-1 excitation of the quantum Hall droplets. In Sec.~\ref{sec:other}, we apply the similar argument to QCD at large isospin density and two-color QCD at large baryon density. Section~\ref{sec:discussion} is devoted to discussions.

In this paper, we use the spacetime metric $g_{\mu\nu} = \diag (+1, -1,-1,-1)$.

\section{Review of two-flavor color superconductivity}
\label{sec:2SC}
We consider massless two-flavor QCD with up and down quarks at finite quark chemical potential $\mu$ at zero temperature. (Inclusion of small up and down quark masses is straightforward.) Based on the Bardeen-Cooper-Schrieffer (BCS) theory, the Fermi surface of quarks at high density is unstable against the formation of quark-quark pairing by the attractive one-gluon exchange interaction and instanton-induced interaction in the color antisymmetric channel, leading to the two-flavor color superconductivity~(2SC)~\cite{Bailin:1983bm,Alford:1997zt,Rapp:1997zu}. Energetically, the spin-$0$ (s-wave) pairing is favored, as it allows all the quarks near the Fermi surface to participate equally in the pairing. This means that the pairing is antisymmetric in spin. According to the Pauli principle, the pairing must also be antisymmetric in flavor. Then, the diquark condensates for right- and left-handed quarks, $q_{\rm R}$ and $q_{\rm L}$, are expressed as
\begin{align}
\label{diquark}
\vev{(q_{\rm R})^{a}_{i} C (q_{\rm R})^{b}_{j}}= \Phi_{\rm R} \epsilon^{ab3}\epsilon_{ij}\,, \quad
\vev{(q_{\rm L})^{a}_{i} C (q_{\rm L})^{b}_{j}}= \Phi_{\rm L} \epsilon^{ab3}\epsilon_{ij}\,,
\end{align}
where $a, b$ are color indices, $i, j$ are flavor indices, and $C$ is the charge conjugation matrix. The ground-state energy is minimized when $\Phi_{\rm R}$ and $\Phi_{\rm L}$ satisfy $\Phi_{\rm R} = - \Phi_{\rm L}$ due to the instanton-induced interaction. 

In the 2SC phase, one of three colors does not participate in pairing, which we choose to be blue with the index $a = 3$ in \er{diquark}. The $\SU(3)_{\rm c}$ gauge symmetry is then Higgsed to ${\SU(2)_{\rm c}}$, resulting in five massive gluons. 
The notable feature of the 2SC phase is that it does not break chiral symmetry and $\U(1)$ baryon number symmetry \cite{Alford:1997zt}.%
\footnote{In this paper, we assume the absence of the chiral condensate in the 2SC phase at sufficiently high density. Physically, the energy cost to excite an antiquark from the Dirac sea becomes larger at higher density, so the chiral condensate is disfavored. Also, while in the color-flavor locked (CFL) phase \cite{Alford:1998mk}, the instanton-induced interaction in the presence of the diquark condensate generates a chiral condensate~\cite{Hatsuda:2006ps,Yamamoto:2007ah}, such a mechanism is absent in the 2SC phase.}
The diquark condensate (\ref{diquark}) is a singlet under the $\SU(2)_{\rm L} \times \SU(2)_{\rm R}$ chiral symmetry. Also, the condensate is a singlet under the $\U(1)_{\tilde {\rm B}}$ modified baryon number symmetry, whose generator is a linear combination of the original baryon number $B = \frac{1}{3}{\rm diag}(1,1,1)$ and the broken color generator $t_8 = \frac{1}{2\sqrt{3}}{\rm diag}(1, 1, -2)$ (which is one of the generators of the coset $\SU(3)_{\rm c}/\SU(2)_{\rm c}$),
\begin{equation}
\label{Btilde}
\tilde B = B - \frac{2}{\sqrt{3}}t_8 = {\rm diag}(0,0,1)\,.
\end{equation}
This shows that the modified baryon number $\tilde B$ is carried only by unpaired (blue) quarks.

\subsection{Confined $\SU(2)_{\rm c}$ gluons}
\label{sec:confinement}
The low-energy degrees of freedom well below the gap $\Delta$ are massless $\SU(2)_{\rm c}$ gluons, unpaired blue quarks, and a pseudo-Nambu-Goldstone mode to be explained later. As the $\SU(2)_{\rm c}$ gluons do not interact with the unpaired blue quarks, the low-energy effective theory for gluons in the 2SC phase is a pure Yang-Mills theory~\cite{Rischke:2000cn}. Note that the Lorentz invariance is explicitly broken by the presence of the medium.

The action of this effective theory can be constructed based on the $\rm SU(2)_{\rm c}$ gauge symmetry, rotational symmetry, and parity symmetry as 
\begin{align}
\label{S_YM}
S=\frac{1}{g^2}\int\rd^{4}x \lb(\frac{\epsilon}{2}{\bm E}^{a}\cdot {\bm E}^{a}-\frac{1}{2\lambda}{\bm B}^{a}\cdot{\bm B}^{a}\rb)\,,
\end{align}
where $g$ is the QCD coupling constant, $E^{a}_{i}:= F^{a}_{0i}$ and $B^{a}_{i}:=-\frac{1}{2}\epsilon_{ijk}F^{a}_{jk}$ are color electric and magnetic fields, respectively, and $\epsilon$ and $\lambda$ are parameters that may be interpreted as the dielectric constant and magnetic permeability. Here, $F^{a}_{\mu\nu}~(a=1,2,3)$ is defined as $F_{\mu \nu} = (F_{\mu \nu}^a) t^a$, where 
$F_{\mu\nu} = \d_{\mu} A_{\nu} - \d_{\nu} A_{\mu} + {\rm i}[A_{\mu}, A_{\nu}]$ is the field strength of the $\rm SU(2)_{\rm c}$ gauge fields, and $t^a$ is the $\SU(2)$ generators satisfying $\tr(t^a t^b) = \frac{1}{2}\delta^{ab}$. 

The parameters $\epsilon$ and $\lambda$ can be computed by the weak-coupling computation as~\cite{Rischke:2000qz,Rischke:2000cn}
\begin{align}
\epsilon=1+\frac{g^2 \mu^2}{18\pi^2\Delta^2} \gg 1\,, \quad \lambda = 1,
\end{align}
where $\Delta$ is the BCS gap parameter, which can also be computed as ~\cite{Son:1998uk,Schafer:1999jg,Pisarski:1999tv,Brown:1999yd,Wang:2001aq}
\begin{align}
\label{gap}
\Delta=512\pi^4\exp\lb(-\frac{\pi^2+4}{8}\rb)\mu g^{-5}\exp\lb(-\frac{3\pi^2}{\sr{2}g}\rb)\,.
\end{align}

We can rewrite the above action in a manifestly Lorentz-covariant form by redefining $(x^{0})':= \epsilon^{-1/2}{x^0}$, $(A^{a}_{0})':= \epsilon^{1/2}A^a_{0}$, and $g':= \epsilon^{-1/4}g$ as
\begin{align}
\label{S}
S=-\frac{1}{2{g'}^2}\int\rd^{4}x' \tr(F_{\mu\nu}' F'^{\mu\nu})\,.
\end{align}
Hence, the parameter $\alpha_{\rm s} = g^2/(4\pi)$ is effectively reduced as 
\begin{align}
\alpha'_{\rm s}:=\frac{g'^2}{4\pi} \simeq \frac{3}{2\sr{2}}\frac{g\Delta}{\mu}\,.
\end{align}
Physically, the diquark condensate is $\SU(2)_{\rm c}$ neutral but screens the $\SU(2)_{\rm c}$ charge by polarization effects, and hence, $\epsilon > 1$. On the other hand, the magnetic fields are not screened at leading order, and $\lambda=1$.

This pure Yang-Mills theory (\ref{S}) has a confinement scale $\Lambda'_{\rm QCD}$, which is different from the usual QCD scale $\Lambda_{\rm QCD}$ in the vacuum. Using the one-loop $\beta$ function, one obtains~\cite{Rischke:2000cn}
\begin{align}
\label{Lambda'}
\Lambda'_{\rm QCD}
\sim \Delta \exp \left(-\frac{3\pi}{11\alpha_{\rm s}'}\right)
\simeq \Delta \exp \left(-\frac{2\sqrt{2}\pi}{11}\frac{\mu}{g\Delta}\right)\,.
\end{align}

\subsection{Pseudo-Nambu-Goldstone mode $\eta$}
At high density, the $\SU(3)_{\rm c}$ instanton effects are suppressed by the Debye screening of gluons~\cite{Shuryak:1982hk}. At asymptotic high density, % they are negligible,  
$\eta$, which is heavy by the instanton effects in the QCD vacuum,
would be a massless Nambu-Goldstone mode associated with the spontaneous breaking of the $\U(1)_{\rm A}$ symmetry by the diquark condensate (\ref{diquark}). 
Below, we will be interested in the sufficiently (but not asymptotically) high-density regime where the explicit breaking of the $\U(1)_{\rm A}$ symmetry is weak. In such a case, $\eta$ can be regarded as a light pseudo-Nambu-Goldstone mode. 

The $\eta$ field is defined as the phase of $\Sigma := \Phi_{\rm R}^{\dag} \Phi_{\rm L}$, namely, $\Sigma = |\Sigma| {\rm e}^{-{\rm i} \eta}$. The effective theory for $\eta$ is then given by~\cite{Son:2000fh,Son:2001jm}
\begin{align}
\label{L_eta}
{\cal L}_{\eta} = f^2 [(\d_0 \eta)^2 - v^2 (\d_i \eta)^2] - V_{\rm inst}(\eta),
\end{align}
where $f$ is the decay constant and $v$ is the velocity in medium, which can also be computed at high density as~\cite{Beane:2000ms}
\begin{align}
\label{f}
f^2 = \frac{\mu^2}{8\pi^2}\,, \quad v^2 = \frac{1}{3}\,.
\end{align}
The potential $V_{\rm inst}(\eta)$ is induced by instanton effects. Because of the Debye screening of $\SU(3)_{\rm c}$ instantons in medium, the typical instanton size is $\rho \sim \mu^{-1}$~\cite{Shuryak:1982hk}.
This makes the semiclassical dilute instanton gas approximation reliable. As a consequence, the potential is dominated by one-instanton contribution and is given by~\cite{Son:2000fh,Son:2001jm}
\begin{align}
\label{V}
V_{\rm inst}(\eta) = - a \mu^2 \Delta^2 \cos \eta,
\end{align}
where
\begin{align}
\label{a}
a \simeq 5 \times 10^4 \left(\ln \frac{\mu}{\Lambda_{\rm QCD}}\right)^7 \left(\frac{\Lambda_{\rm QCD}}{\mu}\right)^{29/3}\,.
\end{align}
This should be contrasted with the case of the QCD vacuum, where the semiclassical instanton computation breaks down in the infrared regime~\cite{tHooft:1976snw}. The $\SU(2)_{\rm c}$ instantons for the low-energy pure Yang-Mills theory~(\ref{S_YM}) may also contribute to the potential $V_{\rm inst}(\eta)$, but its contribution $\sim (\Lambda'_{\rm QCD})^4$ is negligibly small compared with Eq.~(\ref{V}). 

From Eqs.~(\ref{L_eta}) and (\ref{V}), the mass of $\eta$ is obtained as 
\begin{align}
\label{m_eta}
m_{\eta} = 2\pi \sqrt{a} \Delta, 
\end{align}
where we used Eq.~(\ref{f}). For a sufficiently large chemical potential $\mu \gg \Lambda_{\rm QCD}$, we have $a \ll 1$ and $m_{\eta} \ll \Delta$, justifying that $\eta$ is a light Nambu-Goldstone mode. Note that $m_{\eta} \gg \Lambda'_{\rm QCD}$, so the confined $\SU(2)_{\rm c}$ gluon sector lives in the far lower energy regime.

We also note that $\eta$ does not directly interact with the unpaired (blue) quarks, since $\eta$ is introduced through the diquark condensates $\Phi_{\rm R, L}$ for the red and green quarks. However, $\eta$ can interact with the gluons in the confined $\SU(2)_{\rm c}$ sector, which should also be taken into account on top of Eq.~(\ref{L_eta}). We will come back to this issue in Sec.~\ref{sec:QHE}

\subsection{$\eta$ domain wall and $\eta$ string}
As the effective theory (\ref{L_eta}) is the sine-Gordon model, it admits domain-wall solutions. We take the wall to be parallel to the $xy$ plane and consider the profile function interpolating between $\eta = 0$ at $z = - \infty$ and $\eta = 2\pi$ at $z=\infty$.
Then, the domain-wall solution to the classical equation of motion is given by
\begin{align}
\eta(z) = 4 \tan^{-1} {\re}^{m_{\eta}z/v}\,,
\end{align}
which carries a topological charge
\begin{align}
\label{Q_DW}
 Q_{\rm DW} : = \frac{1}{2\pi}\int_{-\infty}^{\infty}
\rd z \partial_z \eta(z) = 1\,.
\end{align}
The tension of the domain wall is~\cite{Son:2000fh}
\begin{align}
\label{T_DW}
T_{\rm DW} = 8\sqrt{2a} f v \mu \Delta \simeq \frac{4}{\pi} \sqrt{\frac{a}{3}} \mu^2 \Delta\,,
\end{align}
where we used Eq.~(\ref{f}).

So far, we have considered an infinite domain wall. Let us now consider a finite domain wall bounded by a closed circular $\eta$ string where the diquark condensate vanishes and around which $\eta$ changes by $2\pi$. The tension of this global string is~\cite{Son:2000fh,Eto:2022lhu}
\begin{align}
\label{T_string}
T_{\rm string} = 2 \pi f^2 v^2 \ln \frac{\ell}{\ell_{\rm core}} \simeq \frac{\mu^2}{12\pi} \ln \left(\frac{1}{2\pi\sqrt{3a}} \right)\,,
\end{align}
where $\ell_{\rm core}$ is the size of the core of the string, which is given by the ultraviolet cutoff of the effective theory, $\ell_{\rm core} \sim \Delta^{-1}$. We also took $\ell$ to be the thickness of the domain wall, $\ell \sim v/m_{\eta}$.

\section{Quantum Hall liquids in 2SC}
\label{sec:QHE}

From now on, we will be interested in the physics of the energy scale $\sim m_{\eta}$. The point neglected in previous studies~\cite{Son:2000fh,Son:2001jm} on the low-energy effective theory (\ref{L_eta}) is that $\eta$ can also interact with the confined $\SU(2)_{\rm c}$ gluon sector via the topological coupling responsible for the QCD anomaly,
\begin{align}
\label{anomaly}
S_{\rm anom} = \frac{N_{\rm f}}{32\pi^2}\int \rd^4 x \eta \tr(F_{\mu \nu} \tilde F^{\mu \nu}) = \frac{N_{\rm f}}{32\pi^2}\int \rd^4 x' \eta \tr(F'_{\mu \nu} \tilde F'^{\mu \nu})\,,
\end{align}
where we explicitly write the number of flavors, $N_{\rm f}=2$, for later purpose, and $\tilde F^{\mu \nu} = \frac{1}{2}\epsilon_{\mu \nu \alpha \beta}F^{\alpha \beta}$. Note that the prefactor of Eq.~(\ref{anomaly}) differs from the corresponding anomalous $\eta F \tilde F$ coupling in the vacuum by a factor of 2. This is because $\eta$ is a four-quark state ($\bar q \bar q qq$) in the color superconducting phase, unlike the two-quark state ($\bar q q$) in the vacuum, and $\eta$ carries the axial charge twice as large as that in the vacuum~\cite{Son:2001jm}. This difference will be important in the following discussion. 

As we will see, this coupling leads to a topological field theory and quantum Hall state on the domain wall.
Similar topological field theory on the domain wall and its consequences in the context of the QCD vacuum were discussed in Ref.~\cite{Komargodski:2018odf}; see also Refs.~\cite{Gaiotto:2017yup,Gaiotto:2017tne} for related earlier works.
Note also that this argument does not apply to the CFL phase in $N_{\rm c}=N_{\rm f}=3$ QCD, as all the eight gluons become massive by the Higgs mechanism~\cite{Alford:1998mk}.

\subsection{Chern-Simons theory on the $\eta$ domain wall}
Let us now consider the effective theory on the $\eta$ domain wall.
We will use the manifestly Lorentz-covariant form and write by omitting the prime, e.g., $(A_0^a)'$ as $A_0^a$, for notational simplicity below. We will ignore the transverse motion of the wall described by the Nambu-Goto action. 

By performing partial integration in Eq.~(\ref{anomaly}), integrating from one side of the domain wall ($\eta=0$) to the other ($\eta=2\pi$), and using the topological charge $Q_{\rm DW}=1$ in Eq.~(\ref{Q_DW}), we obtain the $\SU(N)_{-1}$ Chern-Simons theory on the ($2+1$)-dimensional $\eta$ domain wall $M_3$:%
\footnote{One might worry that the Chern-Simons theory with $N_{\rm f} = 1$ is not properly quantized. However, this effective theory itself is only valid for $N_{\rm f}=2$, as its emergence is specific to the 2SC phase.}
\begin{gather}
\label{CS_SU}
 S_{\SU(N)_{-1}} [A; M_3]
 = -\fr{N_{\rm f}}{8\pi} \int_{M_3} {\rm CS} (A)\,, 
\nonumber \\
 {\rm CS} (A):=  
\epsilon^{\mu\nu\rho} \tr \left(A_{\mu} \d_\nu A_{\rho} 
- \fr{2{\rm i}}{3} A_{\mu} A_\nu A_{\rho} \right) \rd^3 x \,.
\end{gather}
Here, we focus on the contribution of the topological coupling within the domain wall and ignore the boundary terms that may be considered as the interaction of the wall with the outside.
The rank of the gauge group $\SU(N)$ in the low-energy effective theory is $N=2$. The Chern-Simons 3-form satisfies
\begin{equation}
\epsilon^{\sigma\mu\nu\rho}
\d_\sigma \tr \left(A_{\mu} \d_\nu A_{\rho} 
- \fr{2{\rm i}}{3} A_{\mu} A_\nu A_{\rho} \right)
= \fr{1}{2}\tr F_{\mu\nu} \tilde{F}^{\mu\nu}\,,
\end{equation}
with the normalization
\begin{equation}
\fr{1}{2}\int_{\Omega} \rd^4x \tr F_{\mu\nu} \tilde{F}^{\mu\nu} 
\in 8 \pi^2 \mathbb{Z}\,,
\end{equation}
on a 4-dimensional spin manifold $\Omega$. Note that the reason why we get the $\SU(N)_{-1}$ Chern-Simons theory even for $N_{\rm f}=2$ unlike the case in the QCD vacuum~\cite{Komargodski:2018odf} is because $\eta$ is a four-quark state in the color superconducting phase, as mentioned above.
One can then argue that the $\eta$ domain wall carries an anomalous ${\mathbb Z}_N$ 1-form symmetry~\cite{Gaiotto:2017yup,Gaiotto:2017tne}.

This $\SU(N)_{-1}$ Chern-Simons theory can be mapped to the $\U(1)_{N}$ Chern-Simons theory by the level-rank duality~\cite{Hsin:2016blu}, whose action is
\begin{equation}
\label{CS_U(1)N}
S_{\U(1)_N} [a; M_3]
= \fr{N}{4\pi} \int_{M_3} \rd^3 x \epsilon^{\mu\nu\rho} a_\mu 
\d_\nu a_\rho\,.
\end{equation}
Here, $a_\mu$ is a $\U(1)$ gauge field satisfying the normalization conditions,  
\begin{equation}
\fr{1}{2} \int_{\cal S} f_{\mu \nu} \rd S^{\mu\nu} \in 2\pi \mathbb{Z}\,,
\qquad 
\fr{1}{2} \int_{\Omega} \rd^4 x f_{\mu\nu} \tilde{f}^{\mu\nu}
\in 8 \pi^2 \mathbb{Z}\,,
\end{equation}
where $f_{\mu\nu} = \d_\mu a_\nu - \d_\nu a_\mu$ is the field strength, ${\cal S}$ is a closed 2-dimensional subspace, and $\rd S^{\mu\nu} = - \rd S^{\nu\mu }$ is the surface element.
Indeed, the anomaly for the ${\mathbb Z}_N$ 1-form symmetry is matched by the $\U(1)_{N}$ Chern-Simons theory.

The presence of this topological field theory leads to a cusp singularity at $\eta=\pi$ in the potential for $\eta$ on top of Eq.~(\ref{L_eta}).
This provides an additional contribution to the confining tension on the domain wall as
\begin{align}
\label{T_cusp}
T_{\rm cusp} \sim (\Lambda'_{\rm QCD})^3,
\end{align}
where $\Lambda'_{\rm QCD}$ is the modified QCD scale (\ref{Lambda'}) in the confined $\SU(2)_{\rm c}$ sector. However, in the 2SC phase at high density, this contribution is much smaller than the tension of the domain wall in Eq.~(\ref{T_DW}), $T_{\rm cusp} \ll T_{\rm DW}$. This should be contrasted with the case in the QCD vacuum in the large-$N$ limit where $T_{\rm cusp} \gg T_{\rm DW}$ \cite{Komargodski:2018odf}.

Another important point is the absence of the coupling of this Chern-Simons theory to the baryon number gauge field. This is because, in the 2SC phase, the baryon number is modified as $\tilde B$ in Eq.~(\ref{Btilde}) and is carried only by the unpaired (blue) quarks, to which the $\U(1)_{N}$ Chern-Simons theory (\ref{CS_U(1)N}) does not couple. This is one of the main differences from the case in the QCD vacuum~\cite{Komargodski:2018odf}.

\subsection{Quantum Hall droplet as a vector meson}
Let us now consider the case where the domain wall $M_3$ has a boundary. We put the $\U(1)_{N}$ Chern-Simons theory on a cylinder with the boundary being a circle of radius $R$ in the polar coordinates $(r, \theta)$. 

Following the argument of Ref.~\cite{Wen:1992vi}, we choose the gauge fixing condition $a_t + \omega a_{\theta} = 0$, where $\omega$ is a parameter that will be identified as an angular velocity of the edge mode later. We define the new coordinates $\tilde \theta := \theta - \omega t$, $\tilde t := t$, $\tilde r := r$, and then the gauge fields in these coordinates are $\tilde a_{\tilde t} := a_t + \omega a_{\theta} = 0$, $\tilde a_{\tilde \theta} = a_{\theta}$, $\tilde a_{\tilde r} = a_r$. The action is rewritten as
\begin{equation}
\label{CS'_U(1)N}
S_{\U(1)_N} [a; M_3]
= \fr{N}{4\pi} \int_{M_3} \rd^3 x \epsilon^{{\tilde \mu}{\tilde \nu}{\tilde \rho}} {\tilde a}_{\tilde \mu} 
\d_{\tilde \nu} {\tilde a}_{\tilde \rho} =: S_{\U(1)_N} [\tilde a; M_3]
\,.
\end{equation}
The equation of motion for $\tilde a_{\tilde t}$, $\tilde f_{\tilde \theta \tilde r}=0$, is regarded as a constraint. This means that the gauge field $\tilde a_{\tilde i}$ ($\tilde i = \tilde \theta, \tilde r$) is a pure gauge, and can be expressed as $\tilde a_{\tilde i} = \d_{\tilde i} \phi$, where $\phi$ is a scalar field. By inserting it into Eq.~(\ref{CS'_U(1)N}) and performing integration over $\tilde r$ direction, we obtain the edge action
\begin{equation}
S_{\rm edge} = \frac{N}{4\pi}\int \rd{\tilde t} \rd{\tilde \theta} \d_{\tilde t}\phi\d_{\tilde \theta}\phi = \frac{N}{4\pi}\int \rd{t}\rd{\theta} (\d_t + \omega \d_{\theta})\phi\d_{\theta}\phi\,,
\end{equation}
which is a ($1+1$)-dimensional conformal field theory. We can see that the parameter $\omega$ corresponds to the angular velocity of the chiral edge mode $\phi$.

Let us consider a vertex operator $V_N = {\rm e}^{{\rm i}N \phi}$. 
The spin of this state can be found by considering the two-point function $\langle V_N(z) V_M(0) \rangle$ in the complex coordinate $z = x_1 + {\rm i} x_2$ on the plane. Using $\langle \phi(z) \phi(0) \rangle = -\frac{1}{N} \ln z$, one obtains
\begin{equation}
\langle V_N(z) V_M(0) \rangle \sim {\rm e}^{- NM \langle \phi(z) \phi(0) \rangle} \sim \frac{\delta_{N+M,0}}{z^N}\,.
\end{equation}
This shows that the scaling dimension and spin of the operator $V_N$ is $N/2=1$. As the edge theory has no intrinsic scale, the energy of the edge excitation above the ground state is given by $C/(2\pi R)$ with some positive constant $C$. As we mentioned above, this operator does not carry $\U(1)_{\tilde B}$ baryon number in the 2SC phase unlike the case in the QCD vacuum~\cite{Komargodski:2018odf}.%
\footnote{This edge theory has a $\U(1)$ symmetry (corresponding to the constant shift of $\phi$), which one might identify as the $\U(1)_{\rm B}$ symmetry for red and green quarks in the UV theory. However, the genuine baryon number symmetry in the 2SC phase is $\U(1)_{\tilde B}$, and this $\U(1)$ symmetry, if identified as $\U(1)_{\rm B}$ symmetry, should not have physical significance in the red and green sector.}

This finite-size system can be regarded as a quantum Hall droplet. This droplet has spin $1$ without a baryon number so that we may identify it as an isospin-singlet vector meson in the 2SC phase. The total energy of the droplet with radius $R$ is
\begin{align}
\label{E}
E(R) = \pi R^2 T_{\rm DW} + 2\pi R T_{\rm string} + \frac{C}{2 \pi R}\,,
\end{align}
where $T_{\rm DW}$ and $T_{\rm string}$ are given in Eqs.~(\ref{T_DW}) and (\ref{T_string}), respectively, and the last term is the contribution at the edge of the droplet. Here, we ignored the contribution of $T_{\rm cusp}$ in Eq.~(\ref{T_cusp}), which is much smaller than $T_{\rm DW}$. Without the edge contribution, the energy would be minimized when $R \rightarrow 0$, and the droplet shrinks to zero size. Once the edge contribution is included, the total energy has a minimum at nonzero $R$, implying the presence of a stable droplet. However, this result should be taken with care since the minimum is achieved for $R \sim \mu^{-1}$, which is beyond the applicability of the low-energy effective theory. This is somewhat similar to the situation of the skyrmion as a baryon in the QCD vacuum~\cite{Skyrme:1961vq,Skyrme:1962vh}. In that case, the skyrmion is stabilized by the competition between the kinetic term and the so-called Skyrme term with four derivatives in the chiral perturbation theory, but this competition necessarily violates the applicability of the systematic expansion of the low-energy effective theory.

It is interesting to note that the hadron content of the 2SC phase is similar to that of one-flavor QCD in the vacuum at large $N_{\rm c}$. In both cases, the low-energy hadron is the flavor-singlet meson associated with the $\U(1)_{\rm A}$ symmetry, which would be heavy unless the instanton effect is suppressed by medium effect or by large-$N_{\rm c}$ limit. The above argument suggests that the excited spin-$1$ state of the flavor-singlet channel in the 2SC phase may be realized as the quantum Hall droplet.

\section{Other theories}
\label{sec:other}
The above argument can also be applied to QCD at large isospin density and other QCD-like theories (such as two-color QCD) at large baryon density that do not have a sign problem. 

\subsection{QCD at large isospin density}
Let us first consider QCD at finite isospin density \cite{Son:2000xc,Son:2000by}. Since the argument is essentially similar to the case of the 2SC phase above, we only highlight the main differences. At large isospin chemical potential $\mu_{\rm I}$, the Fermi surface is unstable against the formation of the BCS-type pseudoscalar pairing $\langle \bar u \gamma^5 d \rangle \neq 0$. The BCS gap is given by $\Delta = b |\mu_{\rm I}|g^{-5}{\rm e}^{-3\pi^2/(2g)}$ with some constant $b$. The low-energy dynamics well below $\Delta$ is described by the pure $\SU(3)$ Yang-Mills theory, where the confinement scale $\Lambda'_{{\rm QCD}_{\rm I}}$ is much smaller than $\Delta$. This is different from the case of the 2SC phase (where the emergent low-energy effective theory is the $\SU(2)$ gauge theory), as the BCS pairing $\langle \bar u \gamma^5 d \rangle$ is a color singlet. The flavor-singlet $\eta$ becomes a light pseudo-Nambu-Goldstone mode at sufficiently large density again due to the screening of instantons. In the present case, $\eta$ is a two-quark state because the color-singlet $\langle \bar u \gamma^5 d \rangle$ breaks $\U(1)_{\rm A}$ symmetry spontaneously. Also, the low-energy effective theory for $\eta$ admits domain-wall solutions surrounded by a closed $\eta$ string. Another qualitative difference from the 2SC phase is that the UV theory for the $\eta$ sector couples to the genuine baryon number, and so does the effective theory.

Taking into account these differences, we find that the effective theory on the $\eta$ domain wall is the $\SU(3)_{-2}$ Chern-Simons theory, which is dual to the $\U(2)_3$ Chern-Simons theory coupled to the background baryon gauge field $A^{\rm B}_{\mu}$,
\begin{equation}
\label{CS_U(2)N}
 S_{\U(2)_3} [a; M_3]
 = \int_{M_3} \frac{N_{\rm c}}{4\pi}{\rm CS}(a) + \int_{M_3} \rd^3 x \  \epsilon^{\mu\nu\rho} \fr{1}{2\pi} A^{\rm B}_{\mu} \d_\nu \tr (a_\rho) \,.
\end{equation}
where $N_{\rm c} = 3$ is the number of colors and $a$ is the $\U(2)$ gauge field. The vertex operator for the boundary mode of this Chern-Simons theory is ${\rm e}^{{\rm i}N_{\rm c}\phi}$, whose scaling dimension and spin are both $N_{\rm c}/2=3/2$. Also, the $\U(1)$ symmetry of the edge theory can be identified as the $\U(1)_{\rm B}$ symmetry. As a result, the quantum Hall droplet realized in this case is a spin-$3/2$ baryon. This is somewhat similar to the original scenario for the spin-$N_{\rm c}$/2 baryon in the QCD vacuum at large $N_{\rm c}$ \cite{Komargodski:2018odf}.

\subsection{Two-color QCD at large baryon density}
As another example of the similar physics in sign-free QCD-like theories, we here consider two-color QCD with degenerate two-flavor quarks at large baryon chemical potential $\mu_{\rm B}$ where the BCS-type diquark pairing is realized~\cite{Kanazawa:2009ks}. The diquark condensate in this case is a color singlet and spontaneously breaks $\U(1)_{\rm B}$ and $\U(1)_{\rm A}$ symmetries. (Note that in two-color QCD with degenerate two-flavor quarks at high density, $\SU(2)_{\rm L} \times \SU(2)_{\rm R}$ chiral symmetry is {\it not} broken by the diquark condensate.)
The low-energy dynamics well below the BCS gap $\Delta$ is a pure $\SU(2)$ Yang-Mills theory, with the confinement scale $\Lambda'_{{\rm QCD}_2} \ll \Delta$. The flavor-singlet $\eta$ is a two-quark state and becomes a light pseudo-Nambu-Goldstone mode at sufficiently large density~\cite{Schafer:2002yy}. The additional Nambu-Goldstone mode associated with the $\U(1)_{\rm B}$ symmetry breaking does not couple to $\eta$ at leading order, and it can be ignored for our purpose. 

Repeating the similar argument as above, the effective theory on the $\eta$ domain wall is $\SU(2)_{-2}$ Chern-Simons theory, which is dual to $\U(2)_2$ Chern-Simons theory coupled to the baryon gauge field, given by Eq.~(\ref{CS_U(2)N}) with $N_{\rm c}=2$, via the level-rank duality. The vertex operator for the boundary mode of the Chern-Simons theory is ${\rm e}^{{\rm i} N_{\rm c} \phi}$, with the scaling dimension and spin being $N_{\rm c}/2=1$. In this case, the quantum Hall droplet is a spin-$1$ baryon, which may be seen as an excited state of spin-$0$ diquark baryon in two-color QCD.

\section{Discussions}
\label{sec:discussion}
Our finding would provide a physical realization of the scenario of the quantum Hall droplet~\cite{Komargodski:2018odf} in the color superconducting phase at high density without the large-$N_{\rm c}$ limit.

We here summarize several differences of our setup from Ref.~\cite{Komargodski:2018odf} in the QCD vacuum:
\begin{itemize}
    \item{The flavor-singlet meson $\eta$ in the vacuum is heavy due to the QCD anomaly. One thus resorts to rely on the large-$N_{\rm c}$ limit to describe $\eta$ within the low-energy effective theory~\cite{Witten:1979vv}. 
    On the other hand, the instanton effect is suppressed and $\eta$ is a light Nambu-Goldstone mode in the 2SC phase at high density [see Eqs.~(\ref{a}) and (\ref{m_eta})]. Hence, $\eta$ can be described within the model-independent effective theory even for real $N_{\rm c}=3$ QCD.}
    \item{For the quantum Hall droplet in the vacuum at large $N_{\rm c}$, there are two relevant energy scales $\Lambda_{\rm QCD} \gg m_{\eta}$. On the other hand, there are more hierarchical energy scales in the 2SC phase: $\mu \gg \Delta \gg m_{\eta} \gg \Lambda'_{\rm QCD}$. Note in particular that the emergent energy scale $\Lambda'_{\rm QCD}$ for the confined $\SU(2)_{\rm c}$ sector is much smaller than the other scales due to the medium effect [see Eq.~(\ref{Lambda'})]. Then, the cusp singularity for the $\eta$ potential, which is the dominant contribution in the QCD vacuum, is negligibly small in the 2SC phase [see Eq.~(\ref{T_cusp})].}
    \item{In the 2SC phase, the gluon sector at low energy is the $\SU(2)_{\rm c}$ gauge theory, although the UV theory (QCD) is $\SU(3)_{\rm c}$ gauge theory. This suggests that the quantum Hall droplet there describes a spin-$1$ excitation. Interestingly, this matches the fact that the sector of red and green quarks, in which $\eta$ field is constructed, does not carry genuine baryon number, as argued around Eq.~(\ref{Btilde}). This is why the quantum Hall droplet in the 2SC phase may be identified as a flavor-singlet vector meson. This should be contrasted with the one in the vacuum, which is a spin-$N_{\rm c}/2$ baryon.}
\end{itemize}

To the best of our knowledge, this is the first study of the higher-spin hadrons in the 2SC phase. More generally, hadron spectra and resonances in color superconducting phases have not been fully understood so far. This question would also be important to understand the (un)change of the hadron content and possible phase transition (or continuity) between nuclear matter and color superconducting quark matter. Technically, our argument depends on the particular gauge choice, and it would be interesting to provide a formulation in the gauge-invariant language. We defer these issues to future work.

\section*{Acknowledgements}
The authors thank Y.~Hidaka and Y.~Tanizaki for useful discussions and M.~Eto and M.~Nitta for useful correspondences. 
K.~N. is supported by WPI program ``Sustainability with Knotted Chiral Meta Matter ($\mathrm{SKCM}^2$)'' at Hiroshima University.
N.~Y. is supported by JSPS KAKENHI Grant No.~JP22H01216 and No.~JP24K00631.
R.~Y. is supported by JSPS KAKENHI Grant No.~JP21K13928.

\bibliography{2SC.bib}

\end{document}